\documentclass[prl,showpacs,amsmath,twocolumn,superscriptaddress,letterpaper,floatfix,balancelastpage]{revtex4} 
\usepackage{amssymb,verbatim}
\usepackage{amsmath}
\usepackage{graphicx}
\usepackage[colorlinks]{hyperref} 
\usepackage{units}
\usepackage{bbm}

\newcommand{\jmr}[1]{{#1}}

\newcommand{\ket}[1]{\left | #1 \right \rangle}

\newcommand{\beq}{\begin{equation}}
\newcommand{\eeq}{\end{equation}}
\newcommand{\beqa}{\begin{eqnarray}}
\newcommand{\eeqa}{\end{eqnarray}}

\newcommand{\jo}{\rm J}
\newcommand{\lbl}{\rm L}

\begin{document}
\title{Quantum computational renormalization in the Haldane phase}
\author{Stephen D. Bartlett}
\affiliation{School of Physics, The University of Sydney, Sydney, NSW 2006, Australia}
\author{Gavin K. Brennen}
\affiliation{Centre for Quantum Computer Technology, Macquarie University, Sydney, NSW 2109, Australia}
\author{Akimasa Miyake}
\affiliation{%
\mbox{Perimeter Institute for Theoretical Physics, 31 Caroline Street North,
Waterloo Ontario, N2L 2Y5, Canada} }
\author{Joseph M. Renes}
\affiliation{Institut f{\"u}r Angewandte Physik, Technische Universit{\"a}t Darmstadt, Hochschulstr.\ 4a, 64289 Darmstadt, Germany}
\date{10 September 2010}  

\begin{abstract}
Single-spin measurements on the ground state of an interacting spin lattice can be used to perform a quantum computation.  We show how such measurements can \jmr{mimic renormalization group transformations} and remove the short-ranged variations of the state that can reduce the fidelity of a computation.  This suggests that the quantum computational ability of a spin lattice could be a \emph{robust} property of a quantum phase. We illustrate our idea with the ground state of a rotationally-invariant spin-1 chain, which can serve as a quantum computational wire not only at the Affleck-Kennedy-Lieb-Tasaki point, but within \jmr{the} Haldane phase.

\end{abstract}

\maketitle

Measurement-based quantum computation (MQC) proceeds by performing a sequence of single-spin (local) measurements on an entangled resource state of a lattice or graph.  The canonical example of such a resource is the cluster state~\cite{raussendorf01} on a 2D square lattice, although recently alternatives have been proposed~\cite{fv,nest06,gross07,browne08,barrett09}.  Ideally, such a resource would be \emph{natural}, appearing as the stable ground state of a realistic (experimentally accessible) spin lattice.  It would also be \emph{robust}, insensitive to variations in the parameters of the Hamiltonian, such that its quantum computational ability is attributed to a quantum phase in a similar manner to superconductivity and quantum magnetism.  Evidence of such a quantum computational phase has been suggested in a handful of artificial models~\cite{doherty09,browne08,barrett09}. A central problem \jmr{in} this approach, however, is that short-ranged variations in a phase, irrelevant \jmr{to the low-energy physics}, will in general be extremely deleterious for MQC, where the effect of every single-spin measurement is significant for the computation.
\vspace{-1mm}

In this Letter, we show how local measurements within MQC can transform a ground state in such a way as to physically implement a renormalization group (RG) transformation, identifying such a quantum computational phase and correcting for the short-ranged variations. 
As a specific example, we consider rotationally- and translationally-invariant spin-1 chains, which possess a Haldane phase containing the Affleck-Kennedy-Lieb-Tasaki (AKLT) spin-1  ground state~\cite{AKLT}. \jmr{As shown in~\cite{brennen08}, measurements on the} AKLT state can simulate an arbitrary single-qubit gate sequence in the quantum circuit model, i.e.\ it is a quantum computational wire~\cite{gross07,gross08}, 
and forms a basic constituent of MQC when such chains can be coupled.  We show that ground states within this phase \jmr{can also function as quantum computational wires} by appropriately modifying the \jmr{AKLT} measurement sequences.  This modification can be interpreted as \emph{quantum computationally} simulating a renormalization group \jmr{transformation, distilling} out 
the \jmr{long-range degrees of freedom which are common to the entire} Haldane phase as shown by various \jmr{classical algorithmic} RG methods~\cite{schollwock05,weinstein01,verstraete05, gu09}. 
Although specific to this spin-1 model, our result suggests that a similar technique may be applicable in any phase for which a known resource state is a fixed point of \jmr{an} RG flow. Unlike state filtering
techniques to distill resource states via local measurement~\cite{nest06,gross08,barrett09,chen10},
which are strongly dependent on the precise description of the initial state,
our method implements a \emph{parameter-independent} RG that functions robustly
within the phase.
 
{\it The Haldane phase and logic gates.---}%
The AKLT model was originally proposed to analyze the so-called 1D Haldane phase of a spin-1 chain, which displays several characteristic features (see e.g.~\cite{mikeska04}) such as a spectral gap independent of the system size, a diluted antiferromagnetic order often measured by the string order parameter, and an effective spin-$\frac{1}{2}$ degree of freedom (the edge state) appearing on the boundary of the chain.
A generic translationally- and rotationally-invariant Hamiltonian with nearest-neighbor two-body interactions on a spin-1 chain, which takes the form 
\beq
H(\beta) = J \sum_j \left[{\mathbf S}_{j} \cdot {\mathbf S}_{j+1} 
- \beta ({\mathbf S}_{j} \cdot {\mathbf S}_{j+1})^2 \right],  
\label{eq:H}
\eeq
has a gapped Haldane phase for $J > 0$ and $-1 < \beta < 1$.   At the AKLT point $\beta = - \tfrac{1}{3}$, each term in Eq.~(\ref{eq:H}) is the projector onto the spin-2 subspace of neighboring sites, modulo an additive constant and scale factor.  Moreover, $H(-\tfrac{1}{3})$ is  frustration-free, and its ground states have efficient matrix product state (MPS) descriptions~\cite{MPS}. 

\begin{figure}[htbp]
\hspace{-2.5mm}
\includegraphics[width=8.6cm]{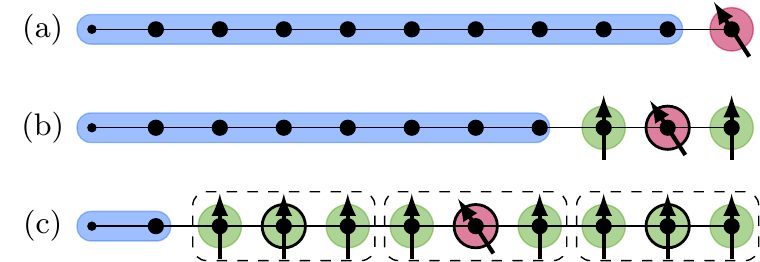}\\
\caption{\label{akltcomp} (Color online.)
Single qubit operations on a Haldane-phase spin chain. A chain of spin-1 particles terminated on the left by a spin-$\frac{1}{2}$ particle encodes one qubit on its right edge. Single spin measurements, shown in (a), implement single qubit operations \jmr{on the AKLT chain}, with measurement in the basis $\{\ket{S_{\hat{e}_i}{=}\,0}\}_{i=1}^3$ for some Cartesian basis $\{\hat{e}_i\}_{i=1}^3$ leading to $\pi$ rotation around the outcome axis~\cite{brennen08}. Each outcome occurs with probability $\nicefrac{1}{3}$. Fixing a ``standard'' basis $\{\hat{x},\hat{y},\hat{z}\}$ (called $\{\ket{x}$,$\ket{y}$,$\ket{z}\}$ in the spin-1 state space), the first two outcomes of measurement in a basis $\{\hat{x}',\hat{y}',\hat{z}\}$ rotated by $\theta$ around the $\hat{z}$ axis (spin-1 states $\{\ket{\theta},\ket{\theta+\pi},\ket{z}\}$, for $\ket{\theta} \equiv \tfrac{1}{2} \left[(1 {+} e^{-i \theta}) \ket{x} + (1 {-} e^{- i\theta}) \ket{y}\right]$) result in the same rotation $R_z(\theta)$ of the qubit, 
followed by a corresponding \emph{byproduct} $\pi$ rotation of it around $\hat{x}$ or $\hat{y}$ (which can be later corrected). The third outcome is just a byproduct $\hat{z}$ rotation. 
Induced rotations become noisy \jmr{for $\beta\neq -\frac{1}{3}$}, but can be improved by \emph{buffering}, depicted in (b). Here the left and right spins are measured first, and the rotation measurement (middle) is only attempted when these are both $\hat{z}$. Failing this, the middle spin is measured in the standard basis, and the attempt is repeated in the next block. Concatenating block-3 buffering is equivalent to block-9 buffering, as shown in (c).
}
\end{figure}

The ground states of finite chains are nearly fourfold degenerate, corresponding to a tensor product of two two-dimensional edge states. 
Each edge state can thus be thought of as a qubit. To associate one qubit to a chain, say on the right, we may fix the left edge by assuming an additional spin-$\tfrac{1}{2}$ particle which terminates the chain, along with an ${\mathbf{s}}\cdot{\mathbf{S}}$ coupling. The encoded Pauli operators then take the form of string operators $\Sigma_k=\sigma_k\otimes e^{i\pi S_k}\otimes\dots\otimes e^{i\pi S_k}$, for $k\in\{x,y,z\}$. This is depicted in Fig.~\ref{akltcomp}, which also details the protocol of~\cite{brennen08} to perform logical operations on the encoded qubit by subjecting the ground state to a sequence of single-spin measurements. 
During the MQC, we do not consider dynamics under  $H(\beta)$; it only specifies a family of resource states. However, controlling $H(\beta)$ at the boundary realizes a quantum computational wire without the RG methods presented here \cite{miyake10}, and may be useful for providing  protection against errors during computation~\cite{brennen08}.
\vspace{-3mm}
\newcommand{\bgthick}{.15}
\def\spspace{.65}
\vspace{-1.1mm}

\begin{figure}[htbp]
\hspace{-2.5mm}
\includegraphics[width=8.8cm]{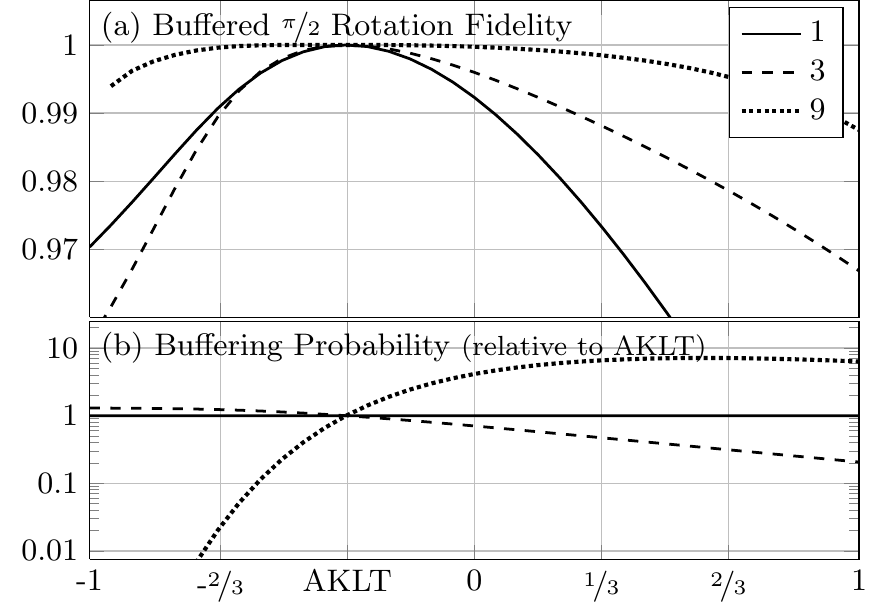}\\
\vspace{-5mm}
\caption{\label{bfid} (a) Fidelity of $\frac{\pi}{2}$ rotation, the worst case, and (b) buffering success probability versus $\beta$ \jmr{for blocklengths $L=1$ (no buffering), 3, and 9} on a chain of length 12. 
The  $L\,{=}\,3$  fidelity decrease for $\beta\,{<}\,-\frac 13$ can be attributed to a one-off effect in our renormalization map, as described in Fig.~\ref{fig:Bloch}.
In (b) the buffering probability is normalized by the AKLT value $\nicefrac{1}{3^L}$; the $L\,{=}\,9$ factor decays to roughly $10^{-5}$ as $\beta\,{\rightarrow}\,{-}1$, though not plotted explicitly due to space constraints.   
Rotation fidelity is computed by entangling the chain with a ficticious termination spin on the right edge  and calculating the overlap of the measured chain with the ideal output state. The unique, entangled ground state of the doubly-terminated chain can be quickly found using sparse matrix methods. This overlap can, in turn, be evaluated using the expectation of string operators, which for
rotation by $\theta$ about the $\hat{z}$ axis (measurement in the $\{\hat{x}',\hat{y}',\hat{z}\}$ basis), are $\Sigma_{x'}\otimes \sigma_{x}$ and $\Sigma_z\otimes \sigma_z$.
The initial state is an eigenstate of $\Sigma_z\otimes\sigma_z$ and remains so after the measurement, implying the square of the fidelity of the output state with the ideal state is given by $F^2=\frac{1}{2}(1+\langle\Sigma_{x'}\otimes\sigma_{x}\rangle)$. 
}
\vspace{-2mm}
\end{figure}

As the protocol described in Fig.~\ref{akltcomp} works for any Cartesian basis, it is reasonable to apply it to the ground state of any
rotationally-invariant Hamiltonian within the Haldane phase.  Indeed, measuring each spin in the same fixed basis implements the resulting $\pi$ rotations, i.e.\ logical identity operations \jmr{modulo Pauli byproducts}, 
with unit fidelity at any point in the Haldane phase. 
However, as shown in Fig.~\ref{bfid} (blocksize 1), the gate fidelity decays away from the AKLT point when measuring in more than one basis.

Such behavior is not unexpected, owing to the relation between rotation fidelity and string operator expectation values as detailed in the caption to Fig.~\ref{bfid}. $R_z(\theta)$ fidelity is related to expectations of $\Sigma_{x'}\otimes\sigma_{x}$ and $\Sigma_z\otimes\sigma_z$ (for doubly-terminated chains), where $\hat{x}'=\hat{x}\cos\theta+\hat{y}\sin\theta$, and for $\theta\neq 2\pi n$ the former is not guaranteed to take on nonzero values throughout the phase. This formulation using string operators also connects the perfect fidelity of standard basis operations to the infinite localizable entanglement length of any ground state in the phase~\cite{campos05}.
\vspace{-3mm}
\vspace{-1mm}

{\it Short-ranged variations and buffered logical gates.---}%
{The reason for the reduced fidelity can be understood by qualitatively examining the variations in the ground state away from the AKLT point.}
{The left-terminated, length-$N$ AKLT chain has the exact MPS description} $\ket{{\mathcal G}(- \tfrac{1}{3})}\propto\sum_{\{b_j\}}\ket{b_N}\cdots\ket{b_1}\otimes\left(\sigma_{b_N}\otimes\cdots\otimes\sigma_{b_1}\right)\ket{\phi}$, for $\ket{\phi}$ an arbitrary spin-$\frac{1}{2}$ state, $\sigma_k$ the Pauli spin operators (except $\sigma_y{\equiv}\sigma_x\sigma_z$), and $b_j\in\{x,y,z\}$ for $j\in\{1,\dots, N\}$.

Observe that the physical swap operation of the spins $j$ and $j{+}1$ is $\mathcal{S}_{j,j+1} := {\mathbf S}_j \cdot {\mathbf S}_{j+1} + ({\mathbf S}_j \cdot {\mathbf S}_{j+1})^2 - \mathbbm{1}$, so that changing the relative weight $\beta$ in Eq.~(\ref{eq:H}) essentially corresponds to a perturbation by swaps.  Heuristically, the ground state $\ket{{\mathcal G}(\beta)}$ for $\beta\approx -\tfrac{1}{3}$ is a coherent superposition of the AKLT state and various partially swapped AKLT states with reordered matrix products.  
Standard basis measurements are unaffected since different orderings of the anticommuting Pauli operators differ at most by a factor $-1$; this alters the probability of the different measurement sequences but not the resulting state. 
However, consider a non-trivial rotation gate $R_z (\theta)$ resulting from measurement outcome $\ket{\theta}$ on spin $j$ followed by $\ket{x}$ on spin $j{+}1$. The dominant AKLT term yields the logical action $ \sigma_x (\sigma_x R_z(\theta))= R_z (\theta)$. But the ordering is reversed in the swapped term, and the action on the logical state is instead $(\sigma_x R_z (\theta))\sigma_x = R_z (-\theta)$.  Interference of the two terms reduces the fidelity of the rotation. 

This picture suggests a variation of the measurement sequences for logical gates which would mitigate the effects of these short-ranged variations (swaps).  Consider a block of three spins, {in which the $R_z (\theta)$ measurement is only attempted} on the middle site \emph{conditioned} on the outcomes $\ket{z}$
on the two neighboring sites.  {The neighboring byproducts $\sigma_z$ commute with $R_z(\theta)$ and thus,} under the action of any permutation within the block, the logical action is invariant (up to a minus sign).

This scheme, called \emph{buffering}, can easily be extended to any odd blocklength $L$. Moreover,  buffered measurements can be concatenated to increase the blocklength; Fig~\ref{akltcomp}(c) depicts three $L\,{=}\,3$ buffered measurements implementing an $L\,{=}\,9$ measurement. Increasing $L$ yields higher gate fidelity, as shown in Fig.~\ref{bfid}, though
this improvement comes at the expense of the (heralded) success probability decreasing exponentially in $L$, or doubly exponentially in the number of concatenated $L\,{=}\,3$ steps. 
\jmr{The buffering overhead is, however, {constant} in terms of the input size of the quantum computation.}

{\it Computational renormalization.---}%
Buffered {gate sequences} are insensitive to the short-ranged variations experienced as one moves away from the AKLT point, {but still utilize the long-ranged degrees of freedom characteristic of the Haldane phase.} 
A renormalization group can be constructed from these degrees of freedom under certain coarse-grainings (e.g.~\cite{weinstein01, gu09}), and therefore one expects that rotation measurements performed on these degrees of freedom, viewed as renormalized spins, will have higher rotation fidelity than measurements performed directly on the physical spins.  The challenge 
in MQC is to perform the appropriate renormalized measurement using only single-site measurements and postselection; for the 1D Haldane chain we {are guided by} the heuristic swap analysis. 

Let us elucidate the relation of $L{=}3$ buffering to the RG.  First, we define the map from three spins to one which generates the RG flow. Then we explain how buffering mimics the desired measurement on the renormalized spin.

The state of three spin-1 particles can be expressed in terms of the total angular momentum of all three spins, with $J{=}0,1,2,3$ components.
Due to the symmetry of the buffering procedure, it will only take notice of components which are invariant under interchange of the first and last spins. This leaves two $J{=}1$ representations, one $J{=}2$, and the $J{=}3$. It is convenient to think of the remaining $J{=}1$ sector as a tensor product $\mathcal{H}_{\jo} \otimes \mathcal{H}_{\lbl}$, where $\mathcal{H}_{\jo}$ carries a $J{=}1$ irreducible representation of SU(2) and $\mathcal{H}_{\lbl}$ is a 2-dimensional `label' subsystem.  An explicit orthonormal basis for this tensor product is
\begin{align*}
  \ket{k}_{\jo}\ket{0}_{\lbl}&=\tfrac{1}{\sqrt{5}}\left(\ket{k}_1\ket{\Psi_0}_{23}+\ket{k}_2\ket{\Psi_0}_{13}+\ket{k}_3\ket{\Psi_0}_{12}\right), \\
  \ket{k}_{\jo} \ket{1}_{\lbl}&=\tfrac{1}{2}\left(\ket{k}_1\ket{\Psi_0}_{23}-2\ket{k}_2\ket{\Psi_0}_{13}+\ket{k}_3\ket{\Psi_0}_{12}\right)\,, 
\end{align*}
where $k\in\{x,y,z\}$ and $\ket{\Psi_0}=\frac{1}{\sqrt{3}}\left(\ket{xx}-\ket{yy}+\ket{zz}\right)$ is the $J{=}0$ state of two spin-1 particles.  Our RG map is then defined as follows:  on each sequential block of three spin-1 particles, we project the state onto the $J{=}1$ sector and subsequently trace out the label subsystem.

\begin{figure}[htbp]
\hspace{-5mm}
\includegraphics[width=8.6cm]{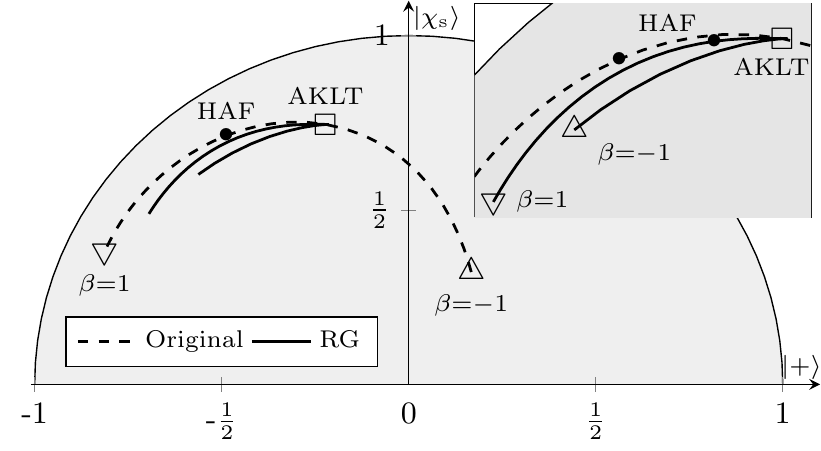}\\
\vspace{-3mm}
\caption{\label{fig:Bloch} Bloch vector of the label space ${\mathcal H}_{\lbl}$ reduced density operator before (dashed) and after (solid) a single RG step for $-1\leq \beta\leq 1$. The state $\ket{\chi_{\rm s}}$ of Eq.~(\ref{eq:chis}) defines the vertical axis, while the horizontal corresponds to the superposition $\ket{+}\equiv\frac{1}{\sqrt{2}}(\ket{\chi_{\rm s}}+\ket{\overline{\chi_{\rm s}}})$, for $\ket{\overline{\chi_{\rm s}}}{=}\frac{1}{\sqrt{6}}(\sqrt{5}\ket{0}{+}\ket{1})$; the $y$ axis is not shown, as the reduced density operator coefficients are real in this basis. 
The Heisenberg antiferromagnet ($\beta=0$) is denoted HAF. 
The norm of the Bloch vector, i.e.\ the radial distance from the origin, provides the weight of
the symmetric three-spin $J{=}1$ sector in the (pre- and post-renormalized) ground state 
$\ket{{\mathcal G} (\beta)}$, 
and the projection onto the vertical axis indicates the buffering success probability, given successful $J{=}1$ projection. 
Observe that for $\beta > -\tfrac{1}{3}$, 
the RG approximately maps Bloch vectors closer to the AKLT point {\it along the original 
curve parametrized by $\beta$}, meaning
the correlation of the reduced state is effectively renormalized to another $\beta$ closer to the AKLT point.
Meanwhile, the first iteration takes $\beta\, {<} -\tfrac{1}{3}$ to $\beta\, {>} - \tfrac{1}{3}$.
Accordingly, we expect that iteration of our RG map generates a flow toward the AKLT point. 
}
\vspace{-3mm}
\end{figure}

This yields a new spin-1 state on $\mathcal{H}_{\jo}$ and, after normalization, a state on a new spin-1 chain of one-third the length. 
Fig.~\ref{fig:Bloch} shows that the map generates an RG flow, depicting the state of the block in the $\mathcal{H}_{\jo}\otimes \mathcal{H}_{\lbl}$ subspace both before and after one iteration of the map. As the system is rotationally-invariant, the state on $\mathcal{H}_{\jo}$ is just the completely mixed state, and we can focus on the Bloch vector of the state on $\mathcal{H}_{\lbl}$. Note that our RG scheme does not map the one-parameter family of ground states exactly onto itself, but nonetheless flows towards states having higher fidelity with the AKLT state.

It is easy to see that the AKLT state is a fixed point of this RG map by using the MPS form.  Projecting two neighboring spins onto $\ket{\Psi_0}$ simply leaves a shorter AKLT chain, since $\sigma_x^2=-\sigma_y^2=\sigma_z^2=\mathbbm{1}$. Similarly, projecting next-neighboring spins onto $\ket{\Psi_0}$ again leaves a shorter AKLT chain, with an overall phase of $-1$. Now consider projecting a block of three spins from $\ket{{\mathcal G}(- \tfrac{1}{3})}$ onto either $\ket{k}_{\jo}\ket{0}_{\lbl}$ or $\ket{k}_{\jo}\ket{1}_{\lbl}$. By the above statements, this yields a shorter AKLT chain, with two spins from the block deleted and one projected onto $\ket{k}$. Tracing out the label subsystem corresponds to mixing these two outcomes incoherently. However, this has no effect, as the two outcomes are identical. Thus, projecting onto the full $\mathcal{H}_{\jo} \otimes \mathcal{H}_{\lbl}$ subspace and tracing out the label subsystem leaves the form of the AKLT state invariant.  

That 3-spin buffered rotation measurement acts as the desired rotation measurement on the renormalized spin arises from the following decomposition of the ``bare'' three-spin states in terms of the tensor product $\mathcal{H}_{\jo} \otimes \mathcal{H}_{\lbl}$,
\begin{align}
  \ket{z,\theta, z}_{123} &\propto \sqrt{\tfrac{2}{5}} 
\ket{\theta}_{\jo} \ket{\chi_{\rm s}}_{\lbl} + J{\neq} 1\ \text{component}\,, \label{eq:chis}\\
  \ket{z,z,z}_{123} &\propto \sqrt{\tfrac{3}{5}} 
\ket{z}_{\jo} \ket{0}_{\lbl} + J{\neq} 1\ \text{component} \,,
\end{align}
where $\ket{\chi_{\rm s}}{=}\frac{1}{\sqrt{6}}(\ket{0}{-}\sqrt{5}\ket{1})$ is \emph{independent} of the measurement angle $\theta$. Buffering implements a projective measurement of the label space, rather than a partial trace, but because $\mathcal{H}_{\jo}$ and $\mathcal{H}_{\lbl}$ are left unentangled, this distinction only affects the success probability and not the resulting map. As the weights of the ``successful'' and ``failure'' outcomes are different, the failure probability of the gate given successful buffering will be different from $\nicefrac 13$ away from the AKLT point; interestingly, it can actually improve, as shown in Fig.~\ref{bfid}(b). In general the $J{\neq} 1$ component of the measurement will have non-zero overlap with the $J{\neq} 1$ component of the state $\ket{\mathcal{G}(\beta)}$, so that the gate fidelity is still less than unity. This is particularly relevant for $\beta\,{\approx}\,{-}1$, where $\ket{\mathcal{G}(\beta)}$ has increased weight on the symmetric three-spin $J{=}2$ subspace.  

{\it Conclusion.---}We have shown that our renormalization protocol removes the
short-ranged variations in the (rotationally-invariant) Haldane phase, generating a flow toward the AKLT point.
Correspondingly, in a practical setting our buffering procedure can be used to ensure a target gate fidelity (chosen by fault-tolerance considerations) for any $|\beta|{<}1$, with attendant decrease in the success probability, as shown in Fig.~\ref{bfid}. In this sense, the quantum computational ability of the spin-chain is a robust property of the phase. 
While we have only given results for a single 1D chain, it is straightforward to include a coupling cphase gate as described in~\cite{brennen08}. Diagonal in the $z$ basis, this gate can thus also be protected by buffering. The resulting fidelity improvement almost exactly follows that of single-qubit rotations as in Fig.~\ref{bfid}(a). 

Our quantum computational RG has several unique features compared with classical RG methods \cite{schollwock05,weinstein01,verstraete05, gu09}.
First, it is a renormalization of a class of states rather than Hamiltonians ~\footnote{Note, however, that our work is 
not based on a short-ranged concatenation of MPS like in~\cite{verstraete05}, even though we use
the MPS description at other points in our analysis.}, though, as  Fig.~\ref{fig:Bloch} shows, it is insensitive to how 
the ``label space'' state is treated.
Remarkably, the map is time-ordered (adaptive in the choices of
later measurements), in contrast with the conventional real-space RG that renormalizes the state or Hamiltonian uniformly in space.
This is crucial to physically implement RG in the \jmr{present context}, \jmr{both because the gates in a quantum circuit provide an implicit time ordering, and due to the need to compensate for the inherent} randomness of measurement outcomes.

Focussing on the
rotationally invariant Hamiltonians of Eq.~(\ref{eq:H}) is well-motivated by physical realizations~\cite{Yip:03}.  For example, for spin-$1$ bosonic atoms trapped in a 1D optical lattice with tunneling-induced interations, the dominant interaction channel is rotationally-invariant $s$-wave scattering.
In realizations using microwave induced dipole-dipole interacting trapped polar molecules, spherical symmetry is provided by the choice of polarization, intensity, and frequency of the applied fields. 

{\it Acknowledgment.---}
SDB and GKB acknowledge the support of the Australian Research Council. 
 SDB, AM, and JMR acknowledge the support of the Perimeter Institute, which receives funding from the Government of Canada through Industry Canada and Ontario-MRI.

\end{document}